\documentclass[a4paper, pra, aps, twocolumn, 10pt, superscriptaddress]{revtex4-2}

\usepackage[T1]{fontenc}
\usepackage{amsmath, amssymb}
\usepackage{empheq}
\usepackage{enumitem}
\usepackage{graphicx}
\usepackage{svg}
\usepackage{booktabs}
\usepackage[braket, qm]{qcircuit}
\usepackage{braket}
\usepackage{hyperref}
\usepackage{placeins}
\usepackage{xcolor}
\usepackage{soul}
\usepackage{tikz}
\usepackage{bbm}
\usepackage{subcaption}

\hypersetup{
	colorlinks   = true, 
	urlcolor     = blue, 
	linkcolor    = blue, 
	citecolor   = red 
}

\begin{document}
\title{Regularizing quantum loss landscapes by noise injection}

\author{Daniil S. Bagaev}\email{bagaev_daniel@list.ru}
\affiliation{National University of Science and Technology ``MISIS”, Moscow 119049, Russia}
\affiliation{Lomonosov Moscow State University, Leninskie Gory 1 building 35, Moscow 119991, Russia}
\author{Maxim A. Gavreev} 
\affiliation{National University of Science and Technology ``MISIS”, Moscow 119049, Russia}
\author{Alena S. Mastiukova}
\affiliation{National University of Science and Technology ``MISIS”, Moscow 119049, Russia}
\author{Aleksey K. Fedorov} \email{akf@rqc.ru}
\affiliation{National University of Science and Technology ``MISIS”, Moscow 119049, Russia}
\author{Nikita A. Nemkov} \email{nnemkov@gmail.com}
\affiliation{National University of Science and Technology ``MISIS”, Moscow 119049, Russia}

\begin{abstract}
The difficulty of training variational quantum algorithms and quantum machine learning models is well established. In particular, quantum loss landscapes are often highly non-convex and dominated by poor local minima. While this renders their training NP-hard in general, efficient heuristics that work well for typical instances may still exist. Here, we propose a protocol that uses a targeted noise injection to smooth and regularize quantum loss landscapes. It works by exponentially suppressing the high-frequency components in the Fourier expansion of the quantum loss function. The protocol can be efficiently implemented both in hardware and in simulations. We observe significant and robust improvements of solution quality across various problem types. Our method can be combined with existing techniques mitigating the local minima, such as the quantum natural gradient optimizer, and adds to the toolbox of methods for optimizing quantum loss functions.
\end{abstract}

\maketitle

\section{Introduction and results}
Variational quantum algorithms (VQA)~\cite{Cerezo2021} and quantum machine learning (QML) models~\cite{Biamonte2017, Schuld2020} are generalizations of classical optimization methods and machine learning (ML) with the potential to harness quantum effects. The range of problems that can be formulated as VQA or QML is extremely broad. However, similarly to classical deep learning, studying VQA and QML remains largely an empirical effort, while firm theoretical performance guarantees are scarce.

One of the key problems facing VQA and QML is trainability. Two main issues here are the onset of barren plateaus (BPs)~\cite{McClean2016, Larocca2024} and proliferation of poor local minima. The BPs essentially manifest the curse of dimensionality, and most often arise in sufficiently deep circuits. Local minima, on the other hand, are typically associated with shallow quantum circuits, which may be free of BPs \cite{Pesah_2021}. (Note, however, that the absence of BPs is often associated with classical simulability \cite{Cerezo2023}.) This work focuses on the problem of local minima in quantum loss landscapes.

On the one hand, results such as \cite{Anschuetz2021, Anschuetz2022, Bittel2021} show that the local minima problem in VQA and QML is intractable in the general case. On the other hand, there may still exist heuristics allowing for efficient training of typical instances despite the no-go results for a general worst-case. After all, training deep neural networks can be very efficient, even though the associated loss functions are typically highly non-convex~\cite{Goodfellow2016}.

\begin{figure}
    \centering
    \includegraphics[width=\linewidth]{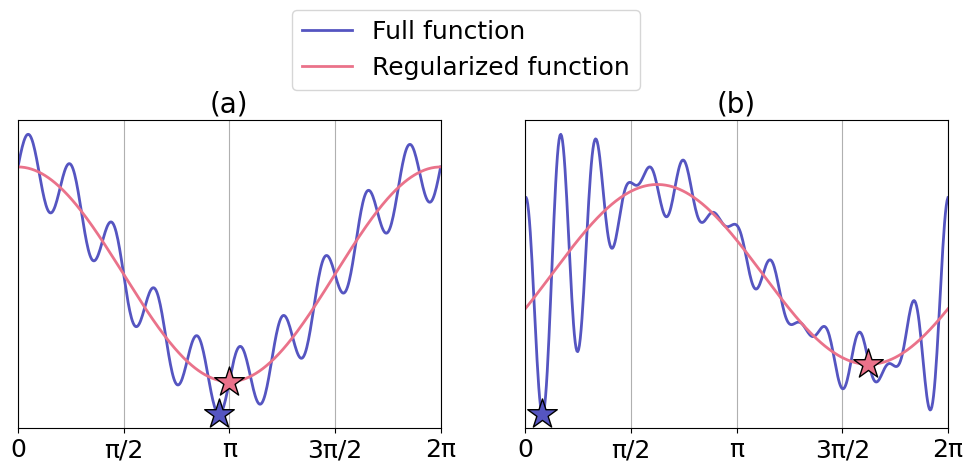}
    \caption{A sketch showing how high-frequency terms may affect the optimization landscape. Stars mark the position of global minima. Both in (a) and (b) the high-frequency terms produce many local minima. However, in (a) the global minimum of the regularized loss function (with high frequency modes discarded) is close to the true global minimum, and can be efficiently used to warm start the optimization. In contrast, in (b) the global minimum of the full function is far from the regularized one.}
    \label{fig sketch}
\end{figure}

In this work, we propose a heuristic regularization procedure that smoothes the quantum loss landscapes and alleviates the problem of local minima. It is based on the intuition that high-frequency terms in the Fourier expansion of the loss function are primarily responsible for the superfluous local minima, see illustration in Fig.~\ref{fig sketch}a. Note that this is merely a sketch and cannot faithfully represent high-dimensional VQA landscapes. Also, Fig.~\ref{fig sketch}b depicts an opposite situation, where the high-frequency modes conspire to produce a global minimum far from the global minimum of the regularized function. Essentially, our conjecture is that the latter scenario is less likely in typical quantum landscapes, and hence regularization of the high-frequency modes is beneficial for optimization.

One analogy to this intuitive picture the use of Gaussian filtering in image processing, which effectively removes high-frequency components from a signal denoising it. Another related concept is that of graduated optimization~\cite{Blake1987}, where the loss landscape during the optimization is first highly smoothed, and then gradually restored to its original form. Graduated optimization have been related to optimization techniques such as simulated annealing \cite{Mobahi2015} and also explored in the context of classical deep learning \cite{Hazan2015, Li2023, Sato2024}. Finally, we note that the Fourier expansion is non-locally related to the original loss function, and can thus give a window into the global properties of the landscape.

\subsection{Results}
In general, the Fourier expansion of quantum loss functions contains exponentially many terms (both in the number of qubits and in the number of parameters; see e.g. \cite{Nemkov2023}), and hence is hard to access or manipulate directly.

The key technical contribution of this work is a protocol that exponentially suppresses high-frequency terms in the Fourier expansion of quantum loss functions. This regularization amounts to injecting a certain amount of noise for each parameterized gate in the quantum circuit, and can be implemented efficiently both in hardware and in simulation. In hardware, a single extra qubit is in principle sufficient to implement the necessary Pauli noise channels, though allowing for more ancillary qubits can reduce the depth of the overall circuit. In software, to simulate our protocol it is sufficient to switch to the density-matrix based simulation, though there is an associated computational cost.

In addition, we show that the regularized and original loss functions are related by the heat equation, which can give both qualitative and quantitative insight to when and why our protocol should make optimization more efficient.

For numerical experiments, we choose two types of models. The first is based on random Wishart fields, which serve as a statistical model for a large class of VQA with local minima \cite{Anschuetz2021, Anschuetz2022}. The second is the quantum convolutional neural network, which is a well-known QML model that has been shown to suffer from local minima already in small circuit sizes \cite{Anschuetz2022}. Our empirical results show a consistent and significant improvement in optimization metrics across all instances studied. For example, the probability to match or improve on the best solution found by the non-regularized optimization typically increases several-fold.

\subsection{Related work}
The problem of local minima in VQA and QML models has long been recognized empirically, see e.g. \cite{Kiani2020, Campos2021, Nemkov2023a}. As the problem is generally NP-hard \cite{Bittel2021}, all existing techniques attempting to solve it are heuristics.

A number of studies systematically explored how the use of different gradient-based and gradient-free optimizers affects the quality of VQA and QML solutions \cite{Lockwood2022, Bonet-Monroig2021}. Notably, quantum natural gradient \cite{Stokes2020} has been claimed to significantly alleviate the problem of local minima in variational quantum eigensolver \cite{Wierichs2020}, though to our knowledge it was only tested in a very limited setting and furthermore incurs significant computational overhead.

Other proposals include using classical heuristics such as differential evolution \cite{Failde2023} and classical neural networks to warm-start the optimization \cite{Zhang2025} or regularize the loss landscape \cite{Rivera-Dean2021}.

Furthermore, the potential benefits of noise to optimization of quantum loss functions have been noted previously. Refs \cite{Nguyen2016, Somogyi2024} argued that quantum noise can improve generalization of quantum neural networks, while \cite{Heyraud2022} made a similar observation for quantum kernel methods.  In \cite{Campos2021} it was observed that a small amount of noise may prevent layer-wise training saturation in quantum approximate optimization algorithms. Ref. \cite{Liu2022b} argued that statistical sampling noise can help avoid saddle points. We also note that our protocol was inspired by work \cite{Fontana2023}, which showed that quantum noise typically exponentially suppresses high-frequency terms in the Fourier expansion of quantum loss functions.

Importantly, our algorithm is complementary to most techniques mentioned above. The use of specific optimizers or warm-starts can be thought of as changing the trajectory of the basic gradient descent to help it avoid local minima and improve solution quality. In contrast, our method alters the loss landscape itself to help smooth or eliminate the superfluous local minima. As such, it can be integrated with most other mitigation methods, such as the quantum natural gradient optimizer.

\section{Theory}
\subsection{VQA and QML}
A VQA is defined by a parameterized quantum circuit $U(\phi)$ (PQC) and a Hamiltonian operator $H$. For simplicity, we assume the initial state to be $\ket{0}$, so that the loss function is
\begin{align}
L(\phi)=\braket{0|U^\dagger(\phi) H U(\phi) |0} \ . \label{VQA loss}
\end{align}

There is a variety of techniques that use PQCs to formulate QML models. We will consider the simplest case of a supervised ML problem, where the data features $x_i$ are encoded into input vectors $\ket{x_i}$ and the model predictions are given by expectation values
\begin{align}
\widehat{y}_i = \braket{x_i|U^\dagger(\phi) H U(\phi)|x_i} \ . \label{predictions}
\end{align}
The loss function is then computed from the ground labels $y_i$ and predictions $\widehat{y}_i$ by standard rules, e.g.
\begin{align}
L(\phi) = \frac{1}{D}\sum_{i=1}^D l(\widehat{y}_i, y_i) \ , \label{QML loss} 
\end{align}
where $D$ is the dataset size and $l$ is a classical loss function (e.g. cross-entropy).

\subsection{Fourier expansion}
We will assume that the PQC $U(\phi)$ consists only of constant Clifford gates $C_k$ and parameterized Pauli rotations $U_P(\phi_k)=e^{i\frac{\phi_k P_k}2}=\cos\frac{\phi_k}2\mathbb{I}+i\sin\frac{\phi_k}{2} P_k$
\begin{align}
    U(\phi) = C_{m+1}\prod_{k=1}^{m} U_{P_k}(\phi_k) C_k  \ .
\end{align}
Virtually all common parameterized quantum circuits studied in modern literature have this form.

For clarity of exposition, we will first focus on the simplest VQA model \eqref{VQA loss}. This loss function admits a natural Fourier-series expansion 
\begin{align}
    L(\phi_k)=\sum_{m=0} L_m(\cos\phi_k, \sin\phi_k) \ . \label{fourier basic}
\end{align}
(Note that the arguments are $\cos\phi_k$ and $\sin\phi_k$, not $\cos\frac{\phi_k}{2}$ and $\sin\frac{\phi_k}{2}$). Here $L_m$ represent all Fourier modes of order $m$. They are homogenous polynomials of degree $m$, meaning that if all arguments of $F_m$ are rescaled by some factor $\lambda$, the function is rescaled by $\lambda^m$ 
\begin{align}
L_m(\lambda \cos\phi_k, \lambda \cos\phi_k) = \lambda^m L_m(\cos\phi_k, \cos\phi_k) \ . \label{homogenuity}
\end{align}

Note that while the number of terms in \eqref{fourier basic} is in general exponentially large, the maximal degree is always bound by the total number of parameters in the circuit, and thus the Fourier series truncates to a Fourier polynomial.

Our main conjecture is that the highly oscillatory higher-order Fourier terms are responsible for the majority of poor local minima, and suppressing them should effectively regularize the loss landscape and increase the quality of solutions. Given this intuition, one approach is to simply discard the high-frequency terms. While straightforward conceptually, this possibility seems non-trivial to implement either in simulations or in hardware. As the number of terms in the Fourier expansion is exponentially large, directly accessing and manipulating each term is not possible beyond small scales.

\subsection{Regularizing Fourier expansion by noise injection   }
Remarkably, there is a protocol that allows to exponentially suppress high-frequency modes, while admitting efficient implementation both in simulation and in hardware. 

The Fourier expansion of the loss function can be computed recursively, using the following simple rule (see e.g. \cite{Nemkov2023})

\begin{align}
U_{P_k}(\phi_k) \circ H_\alpha = \left\{
\begin{array}{@{}r@{\quad}l@{}}
H_\alpha,\hfill & (+) \\
H_\alpha + i \sin{\phi_k} P_k H_\alpha, & (-)
\end{array}
\right. \label{sandwitch}
\end{align}
Here $U_{P_k}(\phi_k) \circ H_\alpha = e^{-\frac{iP_k \phi_k}{2}} H_\alpha e^{\frac{iP_k \phi_k}{2}}$ is the Heisenberg action of $U_{P_k}(\phi_k)$ and  we assume that the Hamiltonian is decomposed into a sum of Pauli strings $H=\sum c_\alpha H_\alpha$, so that each $H_\alpha$ either commutes $(+)$ or anti-commutes $(-)$ with $P_k$. The Fourier expansion of the loss function for the full Hamiltonian $H$ is then simply a sum of the Fourier expansions for individual Pauli terms $H_\alpha$. 

\begin{figure}
\centering
\begin{subfigure}{\linewidth}
\includegraphics[width=\linewidth]{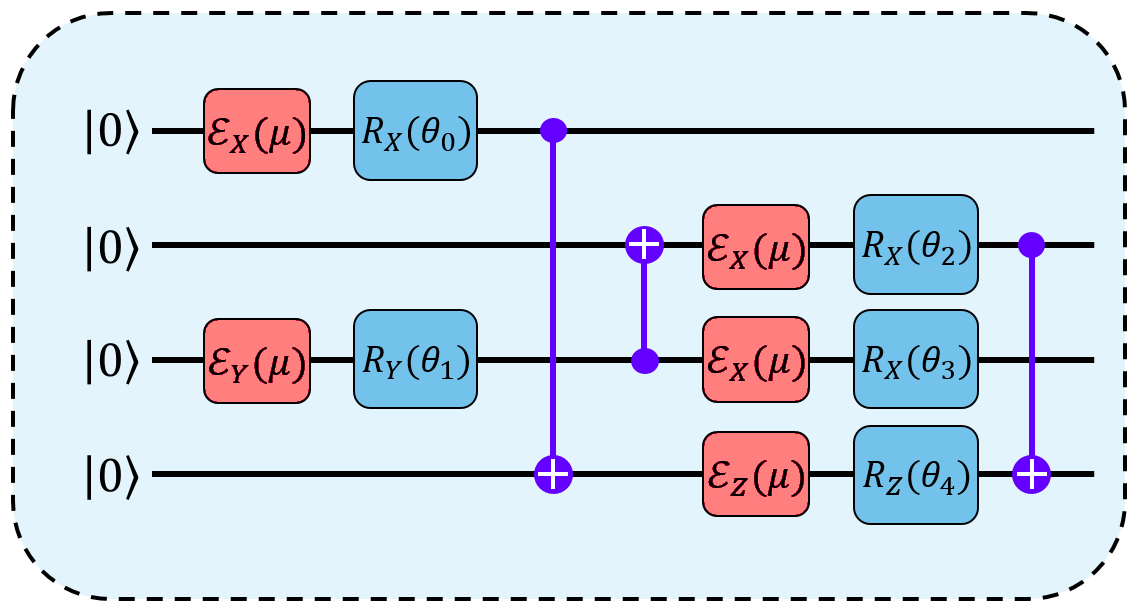}
\caption{A generic quantum circuit modified by injecting the Pauli noise channels according to our protocol.}
\label{fig circuit with noise}
\end{subfigure}

\begin{subfigure}{\linewidth}
\includegraphics[width=0.75\linewidth]{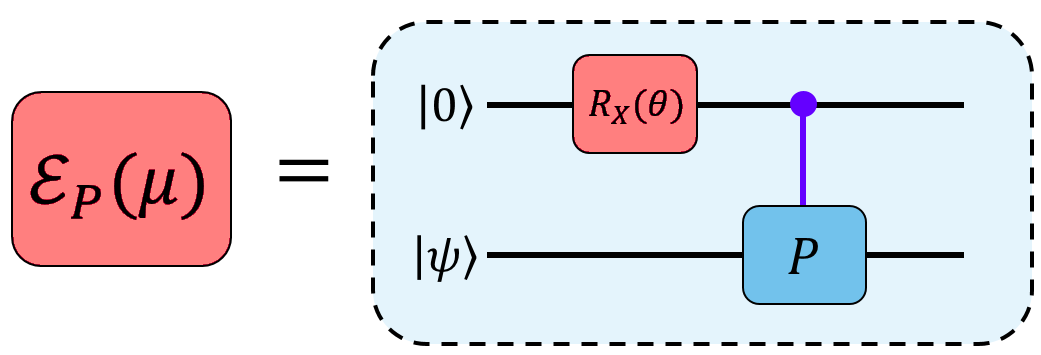}
\caption{An possible implementation of the Pauli noise channel $\mathcal{E}(\mu)$. Here $\mu=2\sin^2{\theta/2}$.}
\label{fig noise channel}
\end{subfigure}
\end{figure}

Now, to a Pauli operator $P$ let us associate a noise channel $\mathcal{E}_P(\mu)$ with Kraus operators $\{\sqrt{1-\frac{\mu}2}\mathbb{I}, \sqrt{\frac{\mu}2}P\}$, which acts in Heisenberg picture as

\begin{align}
C_{P}(\mu)\circ O=\left(1-\frac{\mu}{2}\right) O +\frac{\mu}{2}P O P \ .
\end{align}
Assuming operator $O$ either commutes or anti-commutes with $P$ leads to
\begin{align}
\mathcal{E}_P\circ O= \left\{
\begin{array}{@{}r@{\quad}l@{}}
O, \hfill & O P=+PO \\
(1-\mu)O, & OP=-PO
\end{array}
\right. \ .\label{E cases}
\end{align}
Combining \eqref{E cases} and \eqref{sandwitch} yields
\begin{multline}
\mathcal{E}_{P_k}(\mu)\circ U_{P_k}(\phi_k) \circ H_\alpha= \\\left\{
\begin{array}{@{}r@{\quad}l@{}}
H_\alpha, \hfill & (+) \\
(1-\mu)\left(\cos{\phi_k} H_\alpha + i \sin{\phi_k} P_k H_\alpha\right), & (-)
\end{array}
\right. \ .\label{noise sandwitch}
\end{multline}
Thus, adding to each Pauli rotation $U_P(\phi)$ the associated noise channel $\mathcal{E}_P(\mu)$ effectively rescales $\cos\phi_k$ and $\sin\phi_k$ by $1-\mu$. The homogeneity \eqref{homogenuity} then implies
\begin{align}
L(\mu, \phi)=\sum_{m=0}(1-\mu)^m L_m(\phi) \ . \label{L reg}
\end{align}
Here $L(\mu, \phi)=\braket{0|U(\mu,\phi)^\dagger H U(\mu, \phi)|0}$ is the loss function of the quantum circuit $U(\mu, \phi)$ with the noise channels, see Fig.~\eqref{fig circuit with noise}.

Therefore, this simple noise-injection protocol results in a loss function with exponentially suppressed high-frequency terms, and the regularization factor $1-\mu$ is directly controlled by the noise strength.

\subsection{Implementation}

In principle, a single ancilla qubit is sufficient to implement our noise injection protocol in hardware. Indeed, the Pauli noise channel $\mathcal{E}_P(\mu)$ is equivalent to applying Pauli gate $P$, controlled by an ancilla qubit initialized in the state $|\psi\rangle = R_X(\theta)$ with $\mu=2\sin^2\theta/2$, see Fig.~\ref{fig noise channel}. Immediately after, the ancilla qubit can be reset and reused to implement the subsequent noise channel. Using additional ancilla qubits allows applying several noise channels in parallel. The depth overhead of our protocol depends on how many ancilla qubits are available, as well as how efficient the controlled Pauli gates can be compiled given hardware restrictions. 

Note that so far our discussion assumed that the original quantum circuits are noiseless, and the decoherence only happens due to the controlled entanglement with the ancilla qubits. It may be possible to use the actual quantum noise present in real hardware to achieve a similar effect \cite{Fontana2023, Somogyi2024}.

In software, a density-matrix simulation allows implementing the required noise channels directly. The downside is that the direct density-matrix simulation effectively doubles the system size with the associated increase in computational costs. More efficient approaches, e.g. based on tensor networks \cite{Chen2021, Thompson2025} or Pauli propagation \cite{Nemkov2023, Fontana2023, Begusic2023} may be preferable.

\subsection{Heat equation} \label{sec heat}
There is an interesting physical interpretation of the regularized loss function \eqref{L reg}. Introduce a “time” parameter $t$ so that $1-\mu=e^{-t}$. Then, \eqref{L reg} satisfies the heat equation
\begin{align}
\partial_t L(t, \phi) = \Delta_\phi L(t, \phi) \ .
\end{align}
Moreover, $L(0, \phi)=L(\phi)$ is the original loss function. Hence, if the initial loss function is visualized as a temperature distribution, our regularization is equivalent to letting this temperature distribution to evolve and equilibrate. The stronger the noise, the longer the effective evolution time. In the limit of $t\to\infty$ ($\mu=1$) only the constant term in the Fourier series survives, leading to the flat landscape.

This picture can provide intuitive guidance to when our regularization prescription could be useful. For instance, if the majority of local minima are shallow, and are washed away by thermalization before the interesting solutions are erased, we expect our prescription to be effective.

\section{Numerical experiments}
\subsection{Hyperparameters}
Since our method implies gradient-based optimization over a regularized landscape it requires the usual data such an optimizer, learning rate, number of iterations etc. There is an additional important ingredient, though, which we refer to as the regularization schedule. While at the beginning the landscape is strongly regularized, we need to lift the regularization at some point, since ultimately we are interested in the minima of the original loss. Removing the regularization can be performed gradually, by making the noise strength $\mu$ a function of the iteration number $\mu(i)$, which interpolates between some maximum value $\mu_{\max}$ and $\mu=0$.

A schedule $\mu(i)$ should not fall off to zero too fast, because the optimizer needs to spend enough time in the regularized regime. The schedule should also allow enough steps for the optimizer to navigate the original non-regularized loss landscape, to reach better convergence and solution quality. Finally, the transition between the two regimes should not be too abrupt, to allow the optimizer to track the position of local minima as the landscape is deformed. Through numerical experiments, we found that most schedules satisfying these basic requirements perform similarly, see details in App.~\ref{app schedules}. 

For all experiments reported below, we use the ADAM optimizer \cite{Kingma2015} with learning rate $0.5 \times 10^{-2}$. The initial conditions for multi-start optimization are chosen uniformly at random $\phi\in(0, 2\pi)^{m}$. The noise schedule is that of an exponential decay 
\begin{align}
\mu(i)=\mu_{\max} e^{-a\frac{i} {i_{\max}}}  \label{schedule} \ ,
\end{align}
with $\mu_{\max}=0.9$ and $a=10$. The total number of iterations $i_{\max}$ varies and will be specified for each model. Our code is available at \cite{bagaev}.

\subsection{Toy example}
\begin{figure}
\centering
\includegraphics[width=1.0\linewidth]{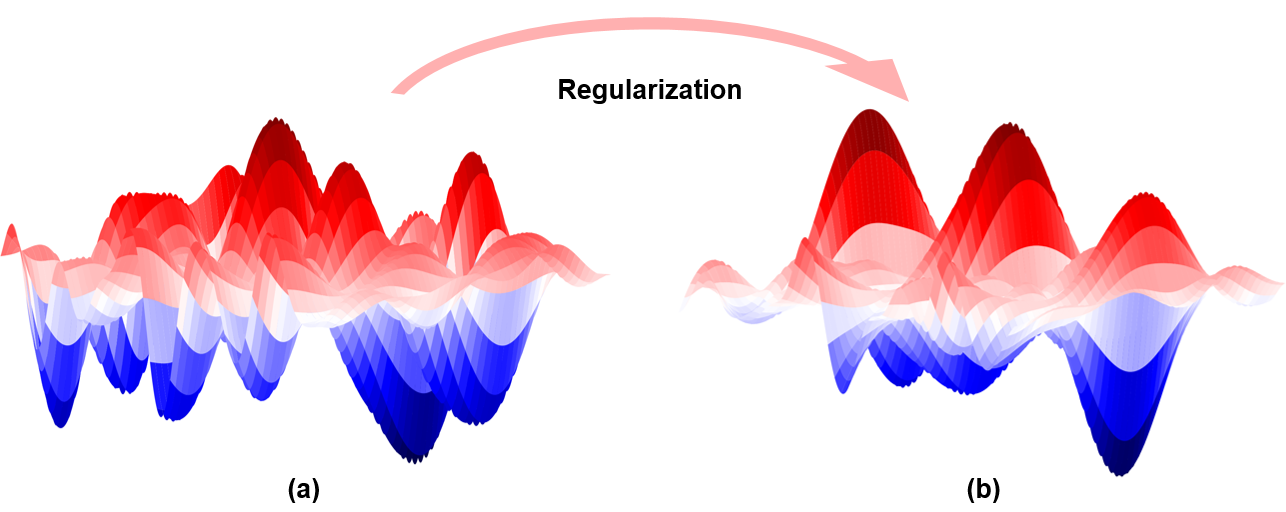}
\caption{A single-layer QAOA loss landscape without (a) and with (b) regularization.}
\label{fig equilibration}
\end{figure}

As a warm-up example, we consider a single-layer QAOA \cite{farhi2014quantumapproximateoptimizationalgorithm} with a 5-qubit Hamiltonian engineered specifically to feature many local minima in its landscape. Since there are only two parameters in the corresponding PQC, the loss landscape can be visualized directly, as shown in Fig.~\ref{fig equilibration}(a). In comparison, Fig.~\ref{fig equilibration}(b) plots the regularized landscape ($\mu=1/3$), which clearly has less pronounced local minima, while keeping the global minimum mostly intact. Loss landscapes are plotted with the help of ORQVIZ \cite{rudolph2021orqvizvisualizinghighdimensionallandscapes} package. We note that the connection between the QAOA data and the Fourier expansion of the loss has been studied in \cite{Stechy2023}.

\begin{figure}
\centering
\includegraphics[width=\linewidth]{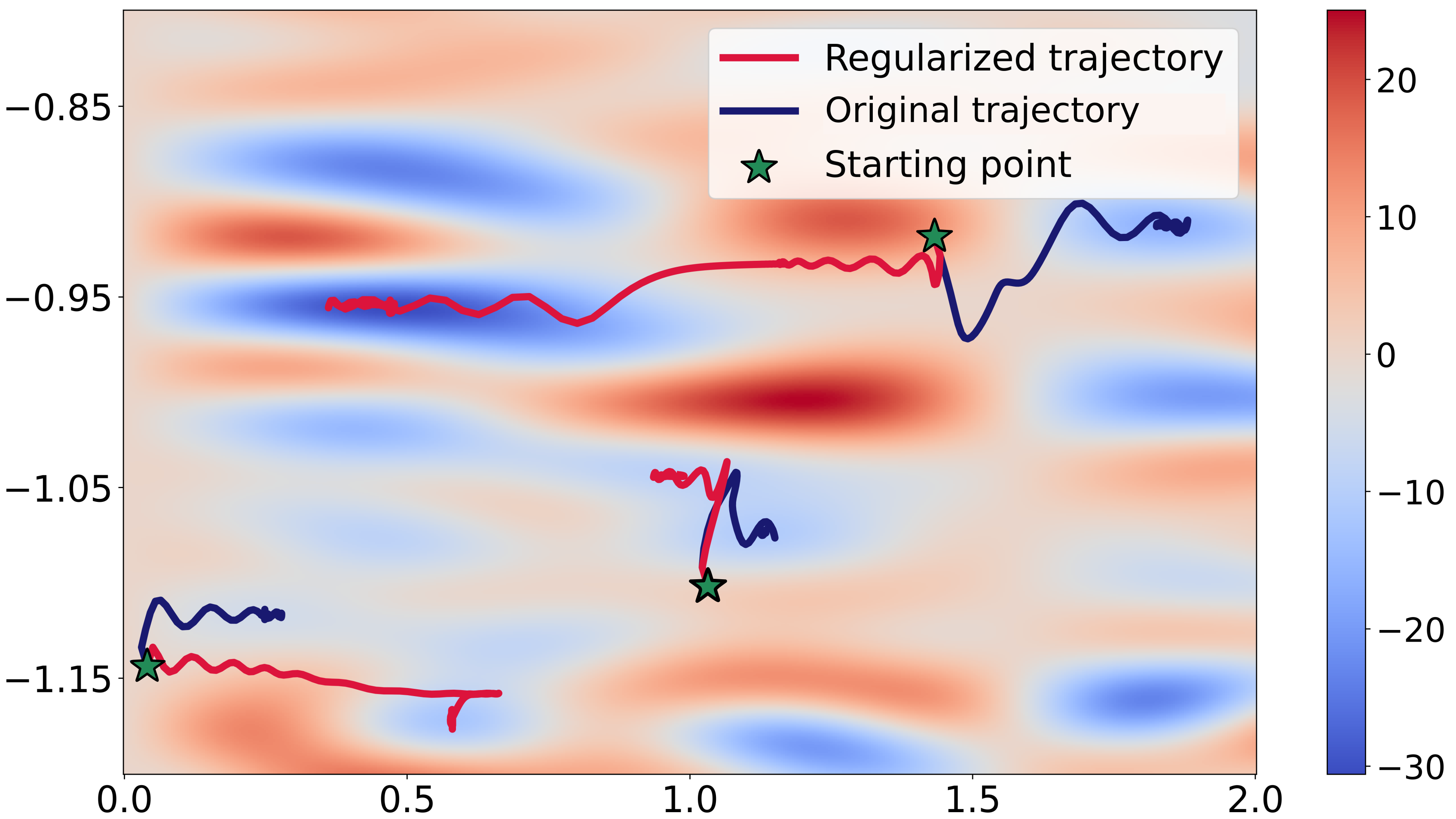}
\caption{Several optimization trajectories traced by the same optimizer navigating the regularized and non-regularized landscapes in our single-layer QAOA example. Only the relevant patch of the landscape is shown.}
\label{fig trajectories}
\end{figure}

Fig.~\ref{fig trajectories} depicts several optimization trajectories of regularized vs. non-regularized optimization starting from the same initial conditions. The background heatmap corresponds to the original loss function. We see that the trajectories traced by the gradient descent in the regularized case are clearly different, and tend to lead to better solutions. The total number of iterations for each trajectory is $i_{\max}=2000$.

\subsection{Statistical model: random Wishart fields}
As shown in \cite{Anschuetz2021, Anschuetz2022}, a large class of quantum landscapes can be described by Wishart hypertoroidal random fields (WHRFs)
\begin{align}
L_{WHRF}(\phi)=w^T(\phi) W w(\phi) \ . \label{L Wishart}
\end{align}

Here $w(\phi)=\otimes_{k=0}^m (\cos\frac{\theta_k}{2}, \sin\frac{\theta_k}{2})$ is a vector of dimension $2^m$, and $W$ is a matrix drawn from a Wishart ensemble, $m$ is the number of parameters in the PQC (see App.~\ref{app Wishart} for details). A Wishart matrix $W$ with $d$ degrees of freedom can be sampled using a Gaussian matrix $X$ with  $2^m$ rows and $d$ columns
\begin{align}
W = \frac{1}{d}X X^\dagger \ . \label{W}
\end{align}
The statistical model \eqref{L Wishart} allows us to study generic features of quantum landscapes, without specifying exact Hamiltonian, PQC structure, or even the number of qubits. For our purposes, the key parameter of WHRFs is the overparametrization ratio
\begin{align}
\gamma = \frac{m}{2d} \ ,
\end{align}
which controls the distribution of local minima \cite{Anschuetz2021, Anschuetz2022}. Namely, for $\gamma \geq 1$ most local minima are close in value to the global minimum, while the opposite is true for $\gamma \ll 1$. Hence, small $\gamma$ is the regime of interest for us here. 

Since the model does not rely on an explicit quantum circuit, our noise-injection protocol does not apply directly. Instead, it is possible to use a different trick achieving the same exponential regularization of the high-frequency modes in the WHRF, see App.~\ref{app Wishart} for details.

To compare optimization over regularized vs non-regularized landscapes, at each value of $\gamma$ we collect extensive statistics by generating 100 different Wishart matrices \eqref{W}, then running 2000 optimizations with $i_{\max}=4000$ iterations, starting from random initial conditions for each landscape. 

\begin{figure}
    \centering
    \includegraphics[width=1.0\linewidth]{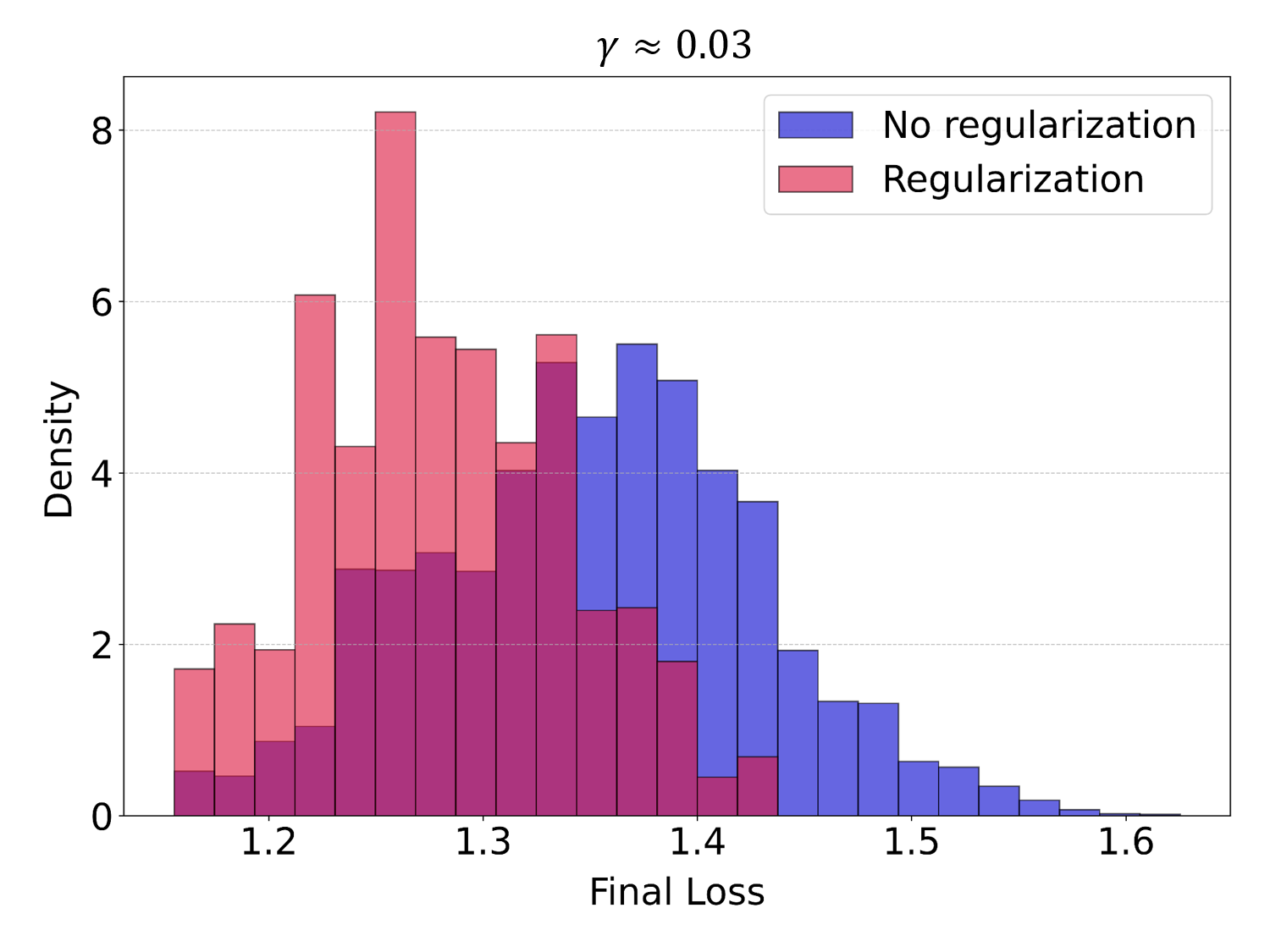}
    \caption{An example pair of histograms showing the distribution of final loss values achieved by regularized and non-regularized optimizations. All loss values correpond to a particular WHRF with $\gamma\approx 0.03$.}
    \label{fig hists}
\end{figure}

Representative histograms of the final loss values are shown in Fig.~\ref{fig hists}. Clearly, optimizations using regularization tend to converge to better solutions on average, and reach the best solutions more often. Note that the true values of the global minima are not available for WHRFs, and by the best solution we mean the best found for a given landscape. 

\begin{figure}
    \centering
    \includegraphics[width=\linewidth]{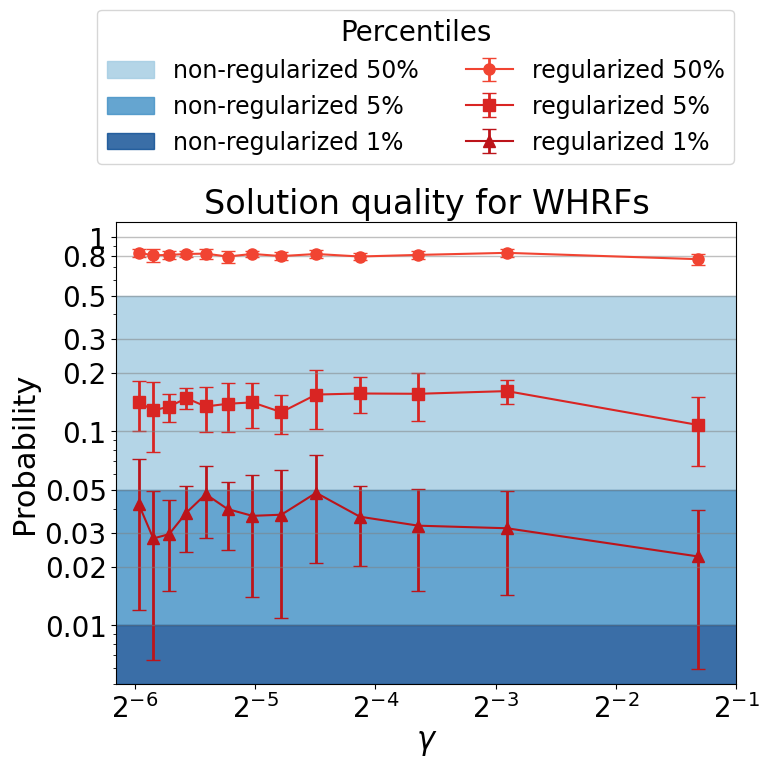}
    \caption{Quality of the regularized optimization as a function of the overparameterization ratio $\gamma$, as measured by the value of $1\%, 5\%$ and $50\%$ percentiles. Background colors correspond the baseline quality obtained by non-regularized optimization. Red lines show percentiles of the regularized value}
    \label{fig wishart}
\end{figure}

To quantify the performance of the regularized optimization systematically, we compare the percentiles of the final loss distributions in Fig.~\ref{fig wishart}. For instance, we compute the first percentile of the non-regularized distribution and see what percent of regularized loss values are below this value. Our show that the probability of finding good solutions (say from the first, or fifth percentile of the non-regularized distribution) is increased 2x to 5x on average, though there is a significant variation across different WHRF instances. Also, we note that this improvement is almost constant as we vary $\gamma$. Also, for sufficiently large $\gamma$ the effect of regularization seems to diminish, probably because the optimization becomes easier for the non-regularized landscapes as well. (Fig.~\ref{fig gammas} from App.~\ref{app Wishart} shows how the quality of non-regularized optimization changes as we cross the $\gamma=1$ threshold).

\subsection{Quantum convolutional neural network}
The quantum convolutional neural network (QCNN) \cite{Cong_2019} is a popular QML model that was shown to be free of BPs \cite{Pesah_2021}.  Although QCNN were recently shown to be classically simulable \cite{Cerezo2023, Bermejo2024}, which undermines their practical utility, they are still an excellent test bed for our optimization technique as they were shown to feature numerous local minima already at small qubit counts.

Another appealing feature of QCNN is that the solution quality can be quantified by a single number -- the accuracy of classification. The training data is generated by the teacher network, an instance of the QCNN with randomly chosen parameters $\phi_*$
\begin{align}
y_i = \langle x_i|U^\dagger(\phi_*) H U(\phi_*)|x_i\rangle \ ,
\end{align}
where the input states $|x_i\rangle$ are random computational basis states. Then, a version of the same network, the student network $U(\phi)$, is trained to replicate the teacher network, starting from an unrelated random parameter configuration. The accuracy is evaluated on the test set, and hence the perfect accuracy can always be reached at $\phi_*=\phi$. Therefore, the train set accuracy directly reflects the difficulty of optimizing the corresponding loss function, which we take to be an MSE $l(\widehat{y}_i, y_i)=\frac{1}{D}\sum_i (\widehat{y}_i- y_i)^2$ following \cite{Anschuetz2022}.

\begin{figure}
\centering
\includegraphics[width=1.0\linewidth]{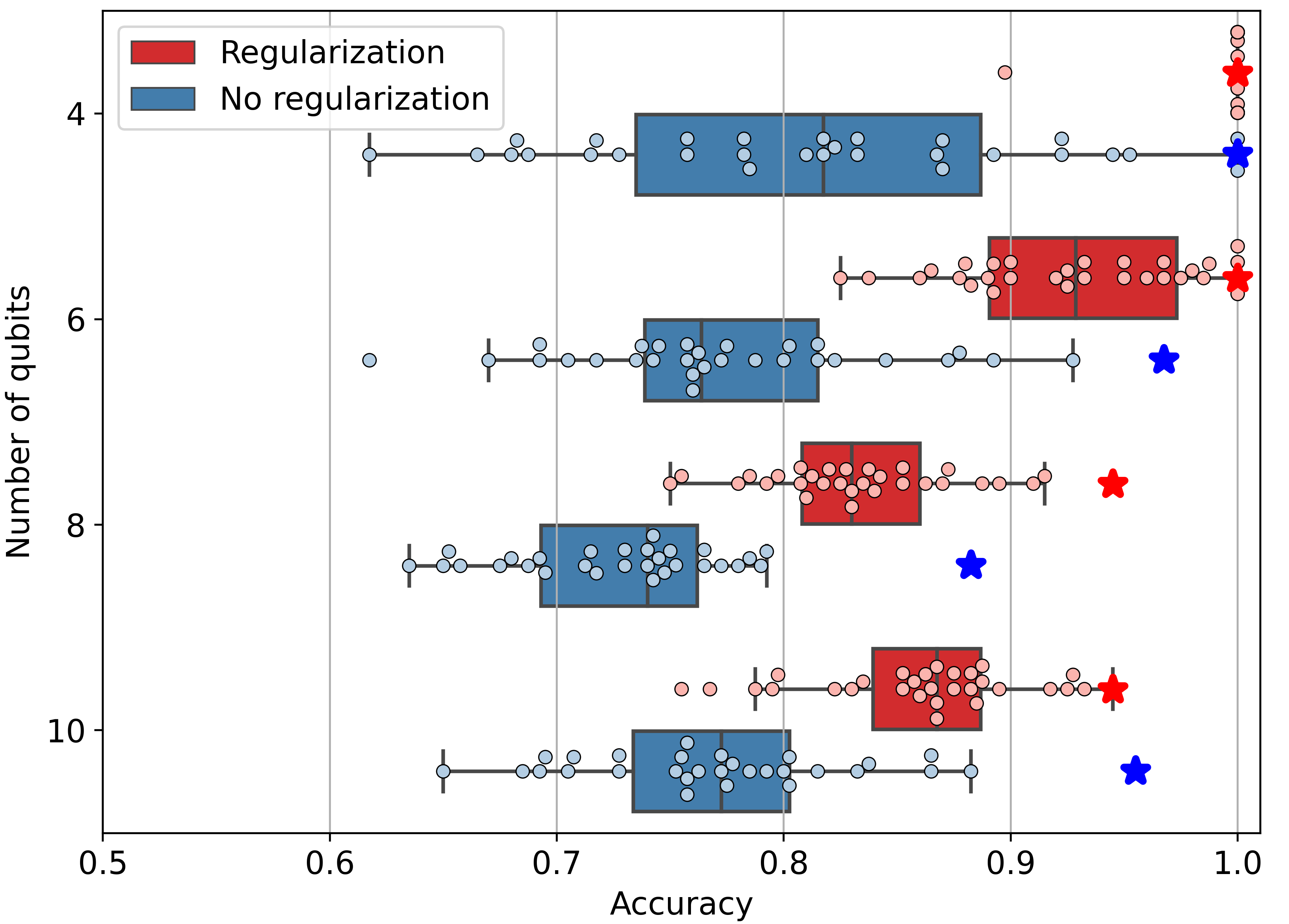}
\caption{Optimization results for QCNN. The highest accuracy values obtained are indicated by stars. To avoid cluttering, the regularized markers are slightly shifted up. In the four-qubit case, the regularized box plot has nearly zero width.}
\label{fig qcnn_boxplots}
\end{figure}

We perform numerical experiments for up to $n=10$ qubits, each featuring a teacher circuit and 30 student circuits, each optimizer over $i_{\max}=2000$ iterations. Results are shown in Fig.~\ref{fig qcnn_boxplots}. Overall, the trend closely mirrors the metrics reported for WRHFs. Both average and best accuracies achieved by the regularized optimization are consistently higher (with the only exception being the best solution for $n=10$ qubits, which is likely an artifact of insufficient statistics). We also note that the results for the non-regularized optimization are consistent with \cite{Anschuetz2022}. Our implementation is an extension of \cite{kiani} supporting noisy quantum channels.

\section{Discussion}
We presented a protocol that uses targeted noise injection to suppress high-frequency Fourier modes, which effectively smoothes and regularizes quantum landscapes. Numerical experiments show that the regularization consistently and significantly improves the quality of optimization solutions. Moreover, as our approach deforms the landscape itself, it is complementary to most other local minima mitigation strategies, and hence can be used jointly with them. However, a number of important caveats need to be addressed to ensure that this technique can be useful in practice.

\subsection*{Noise}

Our procedure assumes the ability to inject variable-strength noise, reducing its value to zero as the optimization proceeds. The actual noise levels on NISQ devices \cite{Preskill2018, Fedorov2022review} may set a threshold that is too high, while the noise mitigation techniques may be too costly \cite{Uvarov2023, Sun2025}. Nevertheless, Ref. \cite{Somogyi2024} suggested that the currently achievable noise levels may already be sufficient to enhance optimization.

Moreover, there are several scenarios where the quantum loss functions are free of the inherent noise. For instance, small-scale circuits which are easily simulable can still have extremely rough loss landscapes, which is a hindrance e.g. to variational compiling \cite{Kiani2020, Nemkov2023a}. In fact, many large-scale VQA and QML loss functions are amenable to efficient classical simulation as well. Some proposals imply that the loss function computation and optimization may be performed entirely classically, while the role of the quantum computer is restricted to the initial data collection or sampling from the final states \cite{Zimboras2025}. Also, VQA and QML are not necessarily a NISQ concept, and may see fault-tolerant implementations in the future. In these cases, the loss landscapes can be assumed noiseless, and our technique should be directly applicable.

It is worth mentioning that the regularization by noise injection may also lead to unwanted side effects. Namely, the noise is known to be one of the causes of barren plateaus \cite{Larocca2024}. For some circuits, the noise strength that provides sufficient regularization of local minima may simultaneously induce barren plateaus. This potential problem is absent in our small-sized numerical experiments, but should be kept in mind when scaling.

\subsection*{Overhead}
Though our regularization robustly improves optimization results, it can not guarantee reaching the desired solution quality every time, and requires an additional resource overhead to implement. Therefore, the extra cost should be compared to e.g. the cost of optimizing the non-regularized loss function multiple times starting with different initial conditions. The results will depend on a particular problem, simulation technique or hardware specs, and we leave a careful analysis of potential trade-offs for future work. We also note that this question is relevant for any other mitigation technique, though rarely discussed explicitly.

Note that in this paper we assumed a very specific noise channel for each of the parameterized gates, which leads to an exact exponential decay of the high-frequency modes. However, it is known \cite{Fontana2023} that quantum noise leads to suppression of the high-frequency modes quite generally. Therefore, different noise injection protocols could result in similar regularization, including those that might be much more efficient to simulate or implement in hardware.

\subsection*{Scaling}
The principle question is whether our mitigation strategy can successfully scale to large problem sizes. Ideally, this should be probed through simulations or actual experiments with practical circuits, which is currently very challenging. Though the small-scale numerical experiments reported here showed that the probability of finding good solutions can be increased several-fold, it is by no means clear that this is sufficient to solve useful problems. After all, the optimization at hand is NP-hard and its difficulty can increase exponentially with the problem size.

We note that the framework of random Wishart fields \cite{Anschuetz2021, Anschuetz2022} may provide a very useful testing ground. The WHRFs are much simpler to simulate at scale, yet should capture the universal behavior of many quantum loss functions. Remarkably, they might even enable some analytic progress, e.g. by studying how the distribution of local minima in WHRFs changes subject to the deformation through heat equation (Sec.~\ref{sec heat}).

Even if it is not possible to overcome the exponential scaling asymptotically, our and other mitigation techniques might still enable solving useful problems more efficiently. However, to the best of our knowledge, no well-defined targets have been established in this space yet; meaning no problem was shown to become feasible to solve if only the probability of finding good solutions could be increased by, say, a hundredfold. Without clear metrics to target, we felt that fine-tuning such as tailoring the learning rate schedule to the regularization schedule is premature. Nevertheless, we expect that the protocol developed in this work will be a useful addition to the quantum loss landscape optimization suite.

\section*{Acknowledgments}

N.A.N. thanks the Russian Science Foundation Grant No. 23-71-01095 (theoretical results). 
Numerical experiments are supported by the Priority 2030 program at the National University of Science and Technology ``MISIS'' under the project K1-2022-027.

\appendix

\section{Schedules} \label{app schedules}
\begin{figure}
    \centering
    \begin{subfigure}{0.49\linewidth}
    \includegraphics[width=\linewidth]{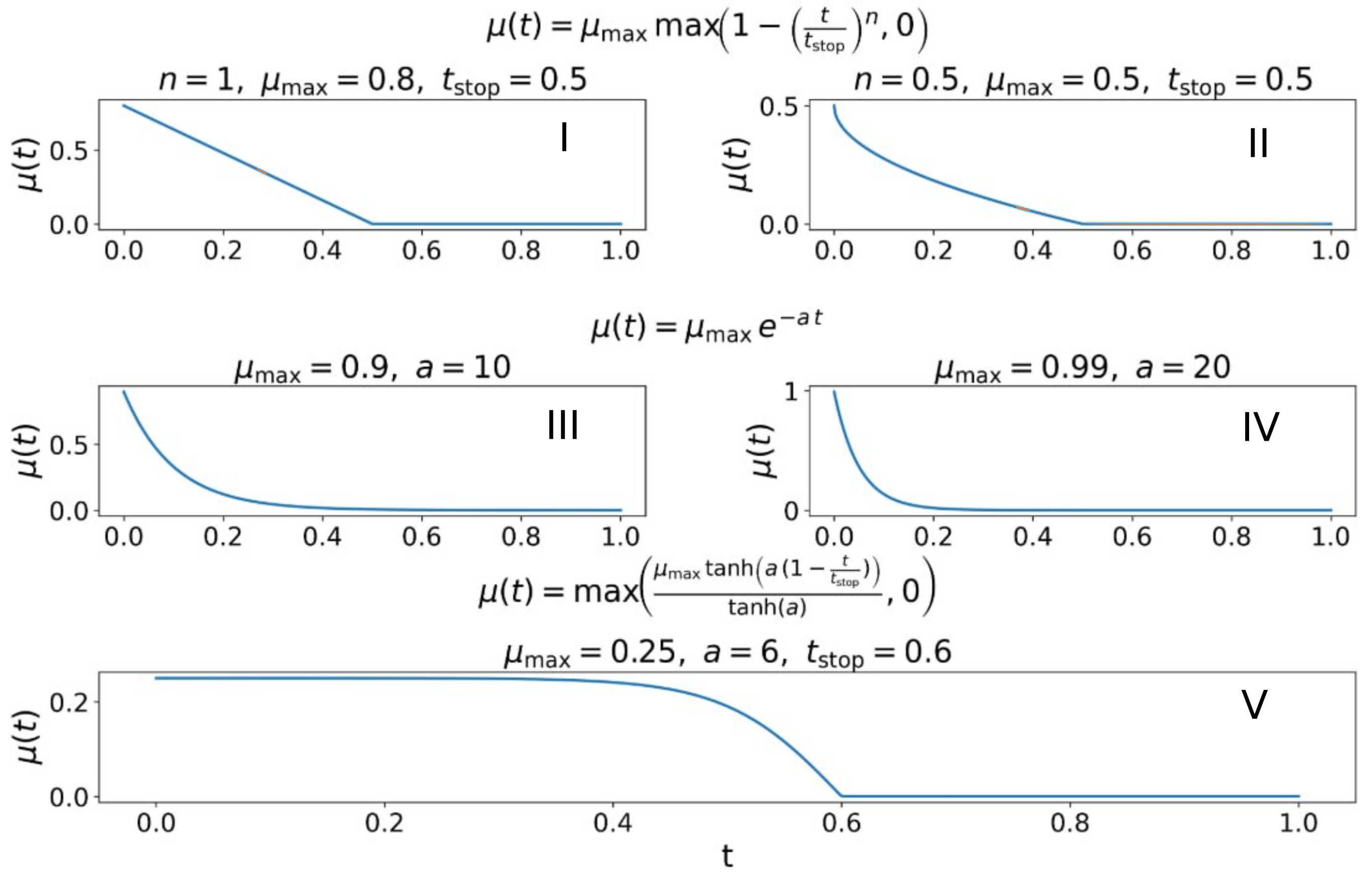}
    \caption{Regularization schedules}
    \label{fig schedules}
    \end{subfigure}
    \begin{subfigure}{0.49\linewidth}
    \includegraphics[width=\linewidth]{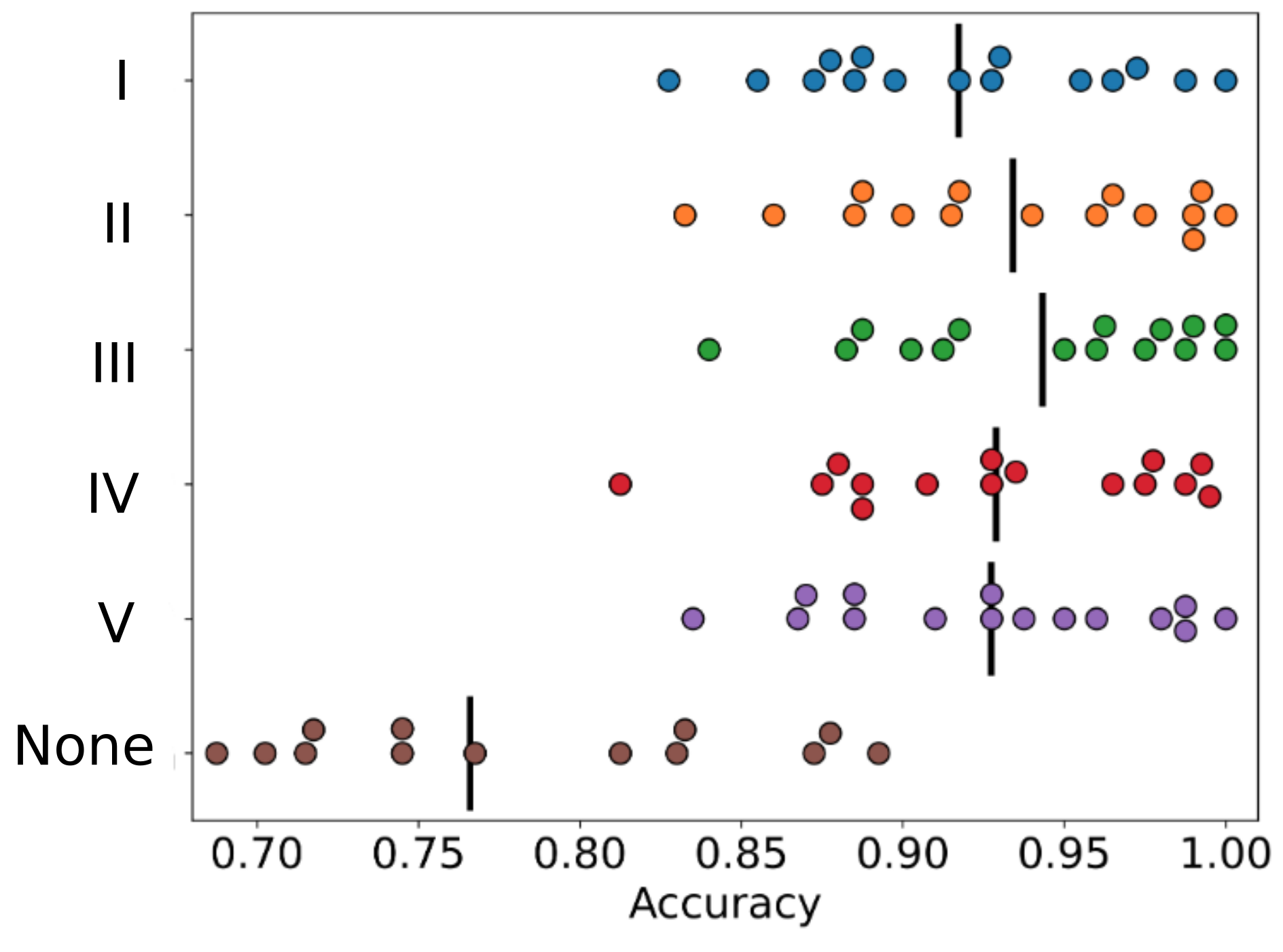}
    \caption{Effect of regularization schedule on accuracy for $n=6$ QCNN model.}
    \label{fig schedules qcnn}
    \end{subfigure}
\end{figure}

We tested several different regularization schedules, some examples are presented in Fig.~\ref{fig schedules}. All schedules seem to perform rather similarly on the instanced we studied, as is illustrated in Fig.~\ref{fig schedules qcnn} for the QCNN case.

\onecolumngrid

~
\twocolumngrid

\section{Wishart random fields} \label{app Wishart}
Loss functions of generic VQAs are expected to be well approximated by Wishart ensembles \cite{Anschuetz2022}
\begin{align}
    L(\phi) = \sum_{i, j} W_{IJ}w_I(\pmb\phi)w_{J}(\pmb\phi) \ . \label{LW}
\end{align}
Here $I, J$ are multi-indices $I=(i_1,\dots, i_p), J=(j_1,\dots, j_p)$. Vectors $w_I(\phi)$ are defined by
\begin{align}
    w_I(\pmb\phi)=\prod_{i_k\in I} w_{i_k}(\phi_k),\qquad w_{i}(\phi) = \begin{cases} \cos{\frac\phi2}, i=0 \\ \sin{\frac\phi2}, i=1 \end{cases}
\end{align}

Exponential suppression of the high-frequency modes in the Fourier expansion of \eqref{LW} can be achieved as follows. Replace
\begin{align}
    w_I(\pmb\phi)w_J(\pmb\phi) \to w_{IJ}(\lambda, \pmb\phi) = \prod_{k}w_{i_k j_k}(\lambda, \phi_k) \ ,
\end{align}
where
\begin{align}
    w_{ij}(\lambda, \phi) = \begin{cases} \frac12(1+\lambda \cos\phi), \quad i=j=0, \\ \frac12 \lambda\sin\phi, \quad i+j=1 \\ \frac12(1-\lambda \cos\phi), \quad  i=j=1 \end{cases} \ .
\end{align}
Note that at $\lambda=1$ we get back the original matrix $w_Iw_J$
\begin{align}
    w_{ij}(\lambda=1,\phi) =\begin{cases} \cos^2\frac\phi2, \quad i=j=0, \\ \sin\frac\phi2 \cos\frac\phi2, \quad i+j=1 \\ \sin^2\frac\phi2, \quad  i=j=1 \end{cases} \ ,
\end{align}
while at generic $\lambda$ the modes of order $m$ are multiplied by $\mu^m=(1-\lambda)^m$.

\begin{figure}
    \centering
    \includegraphics[width=\linewidth]{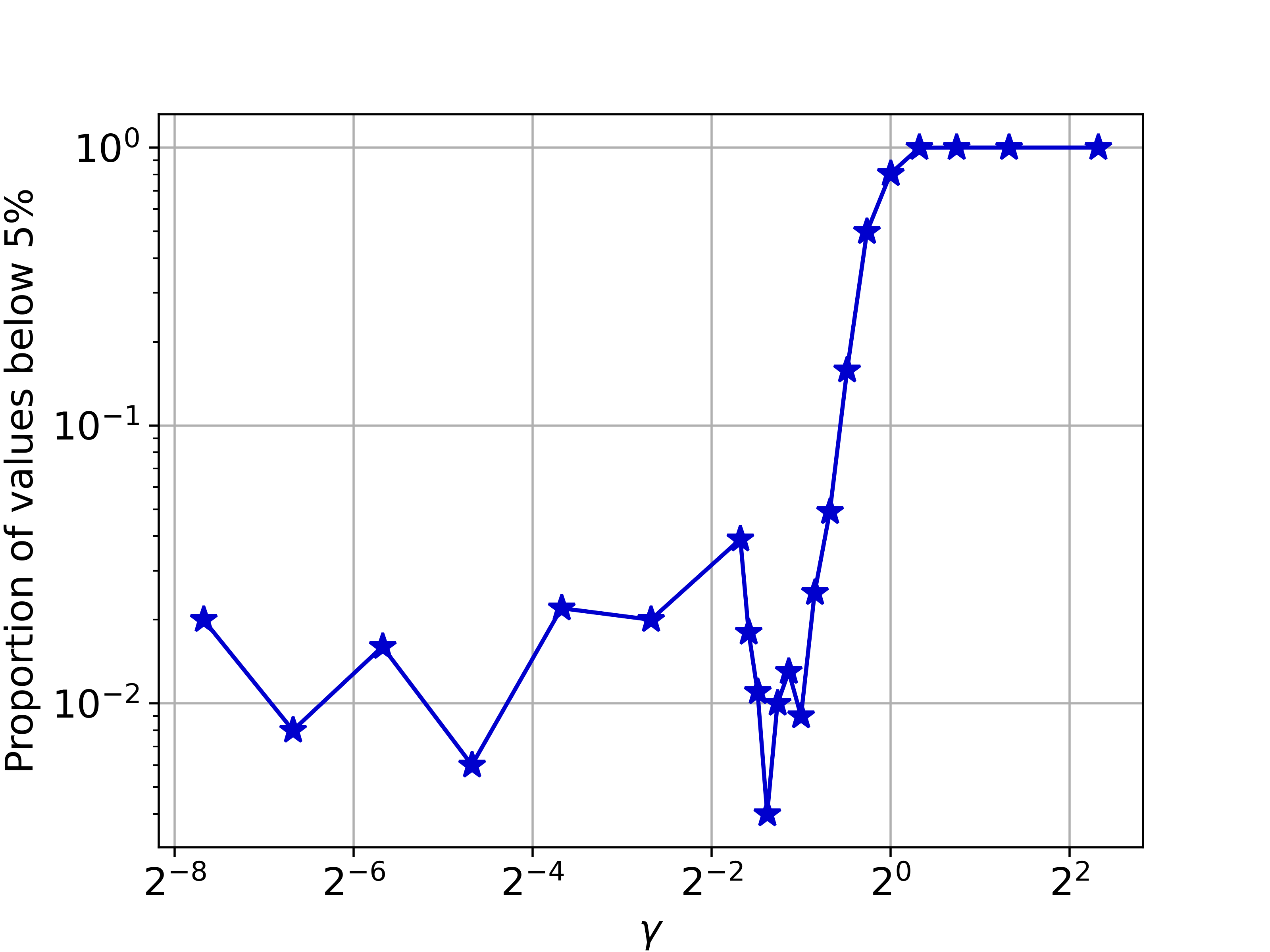}
    \caption{Fraction of loss values in the fifth percentile as a function of $\gamma$ computed under discretization in 100 bins. Past $\gamma=1$ local minima disappear from the landscape.}
    \label{fig gammas}
\end{figure}

\bibliographystyle{quantum}
\bibliography{references}

\end{document}